\documentclass[twocolumn]{emulateapj}

\newcommand{\Lsun}{$L_{\odot}$}

\shorttitle{HD 98800: A 10-Myr-Old Transition Disk}
\shortauthors{Furlan et al.}
\submitted{To appear in ApJ, 2007 August}

\begin{document}

\title{HD 98800: A 10-Myr-Old Transition Disk}


\author{E. Furlan\altaffilmark{1,2}, B. Sargent\altaffilmark{3},
N. Calvet\altaffilmark{4}, W. J. Forrest\altaffilmark{3}, 
P. D'Alessio\altaffilmark{5}, L. Hartmann\altaffilmark{4},
D. M. Watson\altaffilmark{3}, J. D. Green\altaffilmark{3},
J. Najita\altaffilmark{6}, C. H. Chen\altaffilmark{6,7}}

\altaffiltext{1}{NASA Astrobiology Institute, and Department of Physics and 
Astronomy, UCLA, 430 Portola Plaza, Los Angeles, CA 90095; furlan@astro.ucla.edu}
\altaffiltext{2}{NASA Postdoctoral Program Fellow}
\altaffiltext{3}{Department of Physics and Astronomy, University of Rochester, Rochester, 
NY 14627; forrest@pas.rochester.edu, dmw@pas.rochester.edu, bsargent@pas.rochester.edu,
joel@pas.rochester.edu}
\altaffiltext{4}{Department of Astronomy, The University of Michigan, 500 Church St.,
830 Dennison Bldg., Ann Arbor, MI 48109; ncalvet@umich.edu, lhartm@umich.edu}
\altaffiltext{5}{Centro de Radioastronom{\'\i}a y Astrof{\'\i}sica, Universidad Nacional
Aut\'onoma de M\'exico, Apartado Postal 3-72 (Xangari), 58089 Morelia, Michoac\'an, M\'exico; 
p.dalessio@astrosmo.unam.mx}
\altaffiltext{6}{NOAO, 950 North Cherry Avenue, Tucson, AZ 85719;
cchen@noao.edu, najita@noao.edu}
\altaffiltext{7} {{\it Spitzer} Fellow}

\begin{abstract}
We present the mid-infrared spectrum, obtained with the {\it Spitzer} Infrared Spectrograph
(IRS), of HD 98800, a quadruple star system located in the 10-Myr-old TW Hydrae 
association. It has a known mid-infrared excess that arises from a circumbinary disk around
the B components of the system. The IRS spectrum confirms that the disk around 
HD 98800 B displays no excess emission below about 5.5 $\mu$m, implying an optically
thick disk wall at 5.9 AU and an inner, cleared-out region; however, some optically thin dust,
consisting mainly of 3-$\mu$m-sized silicate dust grains, orbits the binary in a ring between
1.5 and 2 AU. The peculiar structure and apparent lack of gas in the HD 98800 B disk suggests 
that this system is likely already at the debris disks stage, with a tidally truncated circumbinary
disk of larger dust particles and an inner, second-generation dust ring, possibly held up by the
resonances of a planet. The unusually large infrared excess can be explained by gravitational 
perturbations of the Aa+Ab pair puffing up the outer dust ring and causing frequent collisions
among the larger particles.
\end{abstract}


\section{Introduction}
During the last few years, particular interest has arisen for a certain
stage in the evolution of circumstellar disks surrounding T Tauri stars:
so-called transition disks, which are characterized by cleared-out or optically
thin inner disk regions and a truncated optically thick outer disk. They
are thought to be in a phase when the disk dissipates rapidly from inside out
\citep[e.g.,][]{strom89, skrutskie90}, on a timescale of $\sim$ 10$^5$ years. 
A few disks in transition have been identified in the 1--2 Myr old Taurus star-forming 
region \citep{forrest04, dalessio05, calvet05}; the different amount of material left 
over in the inner and outer disks of these transition objects implies different formation 
mechanisms or stages in disk evolution \citep{calvet05}.
Planet formation, which is thought to occur in circumstellar disks around T Tauri stars, 
might play a role in the clearing of inner disks \citep{marsh92,calvet02,quillen04,
dalessio05}, but it is not well-known whether other mechanisms, such as photoevaporation,
can produce these transition disks \citep[e.g.,][]{alexander06}. 

On the other hand, disks with inner gaps can also be found in stable, long-lived 
configurations: a binary star system surrounded by a circumbinary disk might create an 
inner gap due to resonant and tidal interactions \citep{artymowicz94}, thus generating 
a spectral energy distribution (SED) typical of transition disks. 
For example, the T Tauri star St 34, which is a spectroscopic binary, has an inner disk 
depleted in dust, likely an effect of the gravitational perturbations of the binary on the 
inner disk regions \citep{hartmann05}. Since this system is likely older than 10 Myr, 
its disk configuration cannot be attributed to a short-lived transitional stage. 

In the $\sim$ 25-member TW Hya association, which is 5--15 Myr old 
\citep{stauffer95, webb99, weintraub00}, four objects are characterized by
significant infrared excesses: TW Hya, Hen 3-600, HD 98800, and HR 4796A
\citep[e.g.,][]{low05}. Except for HR 4796A, whose weaker IR excess clearly
places it into the debris disk category, these stars are surrounded by substantial, 
probably protoplanetary, disks. 
Both TW Hya and Hen 3-600 display close to no disk emission below 7 $\mu$m, an 
indication of truncated inner disks \citep{uchida04}. In particular, TW Hya has been 
suggested as the formation site of a protoplanet based on its inner disk gap 
\citep{calvet02}. Hen 3-600 consists of a spectroscopic binary (the A component)
and a companion separated by 1{\farcs}4 \citep[e.g.,][]{webb99}; the disk
surrounds only Hen 3-600 A \citep{jayawardhana99a}, implying that the inner disk
gap might be a result of gravitational perturbations as in the case of St 34.

In this paper we introduce the mid-infrared spectrum from 5 to 36 $\mu$m of 
HD 98800 (TV Crt) obtained with the Infrared Spectrograph\footnote{The IRS was a 
collaborative venture between Cornell University and Ball Aerospace Corporation 
funded by NASA through the Jet Propulsion Laboratory and the Ames Research 
Center.} \citep[IRS;][]{houck04} on board the {\it Spitzer Space Telescope} 
\citep{werner04}.
HD 98800 is a quadruple system consisting of a visual binary with a projected separation 
of 0{\farcs}8 (which corresponds to 38 AU at the distance of 47 pc determined 
by {\it Hipparcos}) whose components are spectroscopic binaries with separations 
of about 1 AU \citep{boden05}. Mid-infrared imaging revealed that the strong
infrared excess of the system arises from component B only; in addition, it sets in 
at about 7 $\mu$m, implying cleared-out inner disk regions 
\citep{gehrz99,koerner00,prato01}. 
Our mid-infrared spectrum confirms that the disk around HD 98800 is a transition
disk and allows us to determine the location of the circumstellar dust, as well as to 
derive its mineralogical composition.
This paper is structured as follows: in \S\ 2 we present our observations and data 
reduction; in \S\ 3 we construct the SED of HD 98800 B, and fit dust and disk models 
to it; and in \S\ 4 we present a discussion of this transition disk and our conclusions.

\section{Observations and Data Reduction}

HD 98800 was observed during IRS campaigns 9 and 17 on 2004 June 25 and
on 2005 January 03, respectively. 
The observations of campaign 17 repeated those of campaign 9, since 
the observations of the earlier campaign were somewhat compromised by a bright
saturation event. However, the observations of campaign 17 were slightly
mispointed; since we were able to mitigate the effects of the saturation on the array
(apparent mostly in the form of additional rogue pixels), we used the IRS
spectrum of campaign 9 for our analysis.

To obtain the full mid-IR spectrum, we used the Short-Low (SL; 5.2--14 $\mu$m;
$\lambda$/$\Delta\lambda$=60--120),
Short-High (SH; 9.9--19.6 $\mu$m; $\lambda$/$\Delta\lambda$=600), 
and Long-High (LH; 18.7--37.2 $\mu$m; $\lambda$/$\Delta\lambda$=600) modules. 
For a clearer presentation of the IRS spectrum, we rebinned the SH and LH spectra 
to a resolution of 300, and we truncated the SH spectrum below 14 $\mu$m.
The observations were carried out in mapping mode, where we mapped the 
target in 3 steps separated by three-quarters (for SL) or half (for SH and LH) 
of the slit width in the dispersion direction and 2 steps separated by a third of a 
slit length in the spatial direction.

We extracted and calibrated our data with the SMART software tool \citep{higdon04}
after fixing bad pixels in the arrays using a simple interpolation over good, neighboring 
pixels in the spectral direction. In SH and LH, we also fixed all so-called ``rogue'' pixels 
identified in darks from
campaigns 1 to 18 (for LH) or 1 to 28 (for SH). We subtracted the background of
our SL spectrum by using the observation taken in the other nod position in the same
order, then extracted the spectrum using a variable-width column extraction that
scales with the width of the point-spread function. We were not able to subtract any
background in our SH and LH spectra, which we extracted with a full-slit extraction.
The SL spectrum was calibrated with $\alpha$ Lac (A1 V), the SH and LH spectra
with $\xi$ Dra (K2 III), using template spectra from \citet{cohen03}. The final
spectrum was obtained by averaging the spectra of the two central map positions.
Given that the SL, SH, and LH spectra stitched together without applying any scale 
factors, the background contributions in SH and LH are likely negligible.

By comparing the 10 and 24 $\mu$m emission derived from our IRS
spectrum of HD 98800 with ground- and space-based measurements
\citep{jayawardhana99b,low05}, we infer an absolute spectrophotometric 
accuracy of $\sim$ 5\%. Our relative accuracy, determined by the
scattering of neighboring flux values, is higher in SL than SH and LH:
features above the noise level are very likely real in SL, while LH is still
dominated by calibration artifacts.

\section{Analysis}
\subsection{Spectral Energy Distribution}
\label{SED_section}

Previous observations of HD 98800 indicated that the infrared excess originates
from a circumbinary disk around the component B spectroscopic binary.
The SED is consistent with it being a transition disk, extending between a few 
and $\sim$ 10--20 AU \citep{koerner00,prato01}. 
Observations of the H${\alpha}$ line, which did not distinguish the emission from the
four components, revealed a total H${\alpha}$ equivalent width of 0.9 {\AA}
and a width comparable to that of Pleiades K dwarfs, suggesting an origin in
chromospheric activity and not in an accretion flow \citep{soderblom96}. An 
upper limit to the mass accretion rate has not been determined, but it is likely
well below 10$^{-11}$ $M_{\odot}$ yr$^{-1}$ \citep{muzerolle00}; thus, 
accretion onto HD 98800 B has virtually come to a halt.
The absence of accretion signatures \citep{soderblom96, webb99} is in accordance 
with the lack of dust emission from the inner disk, i.e., the outer disk is prevented
from accreting toward the star. 

\begin{figure}
\plotone{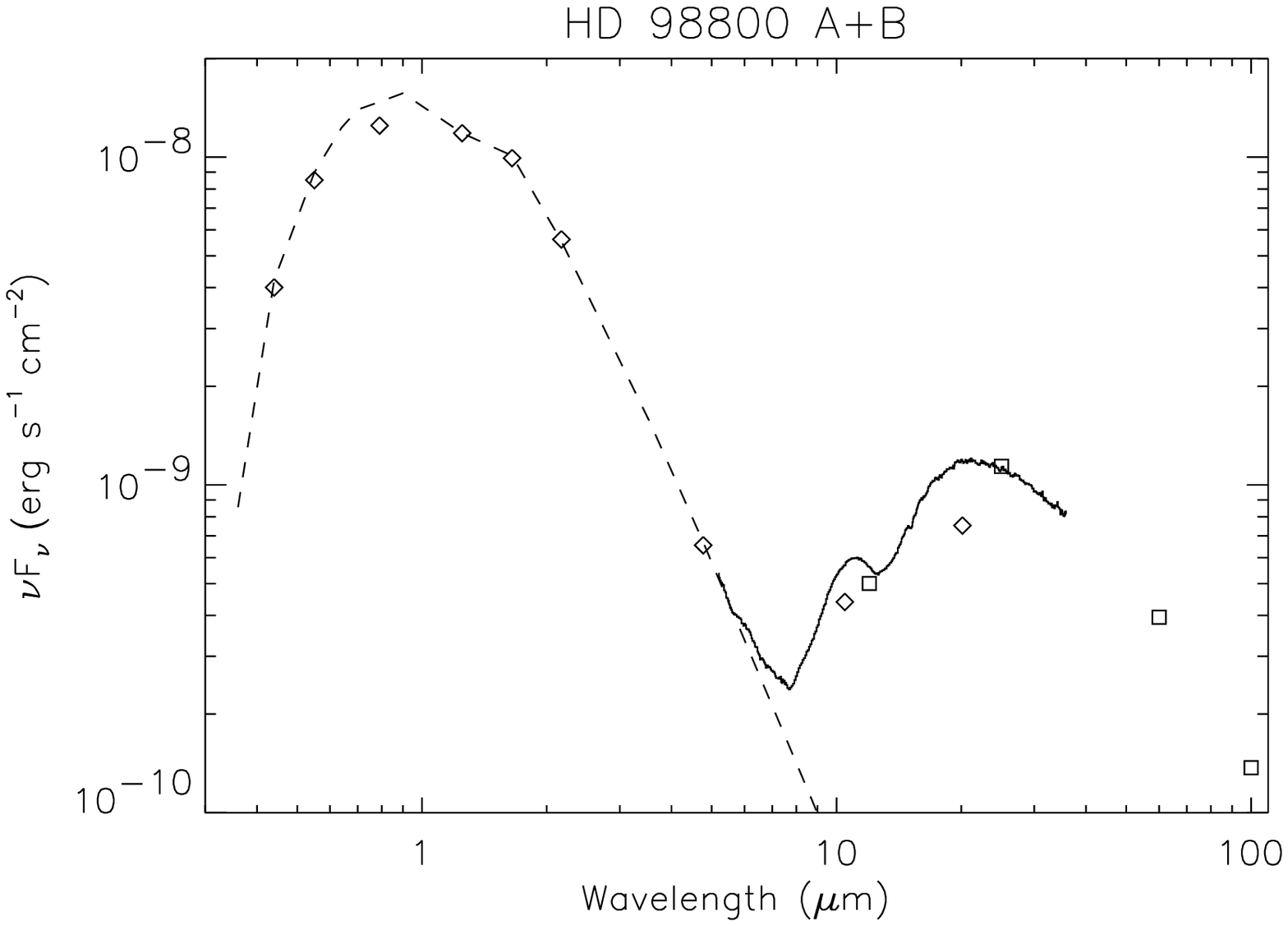}
\caption{SED of HD 98800, composed of the IRS spectrum (which includes 
the flux of all four stars in the system), $B$, $V$, $I$ photometry from 
\citet{soderblom98}, the 2MASS $J$, $H$, and $K_s$ fluxes, 
$M$, $N$, $Q$ photometry from \citet{prato01}, and the {\it IRAS} 
12, 25, 60, and 100 $\mu$m fluxes. The photosphere of a  K5 star (based
on the photospheric colors from \citet{kenyon95} and normalized at $J$) is also 
sketched in. All fluxes were dereddened using $A_V$=0.44 and 
Mathis's reddening law \citep{mathis90}. \label{TWA4_SED}}
\end{figure}
 
Our {\it Spitzer} IRS observations included emission of all four components of 
the system, since the narrowest IRS slit is 3{\farcs}6 wide. Therefore, in the 
SED shown in Figure \ref{TWA4_SED}, we used unresolved optical to far-IR 
photometry of HD 98800 (see figure caption for details) and the IRS spectrum, 
all corrected for reddening using $A_V$=0.44 \citep[e.g.,][]{soderblom98} and 
Mathis's reddening law \citep{mathis90}. Also shown is the photospheric emission
of a K5 star \citep{soderblom98,prato01}, based on the photospheric colors 
of a star with this spectral type \citep{kenyon95} and normalized at the unresolved 
2MASS $J$-band flux. Despite the fact that our IRS observation does not
distinguish the emission from A and B, the SED suggests that none 
of the four components generates excess emission above the level expected 
from a stellar photosphere out to about 6 $\mu$m. This supports the idea 
that the inner disk of HD 98800 has been cleared out.

In order to retrieve the true IRS spectrum of the B component, we estimated
the contribution of the A component and subtracted it from the spectrum.
Component A does not display an infrared excess and has an extinction 
$A_V \sim 0$ \citep{koerner00, prato01, boden05}, while Ba+Bb seem
to lie behind $A_V$=0.44 \citep[e.g.,][]{soderblom98}. This difference in 
extinction likely indicates that we are observing HD 98800 B through 
some circumbinary dust, as first noted by \citet{tokovinin99}; this could also
explain the slight photometric variability observed by \citet{soderblom98}.

\begin{figure}
\plotone{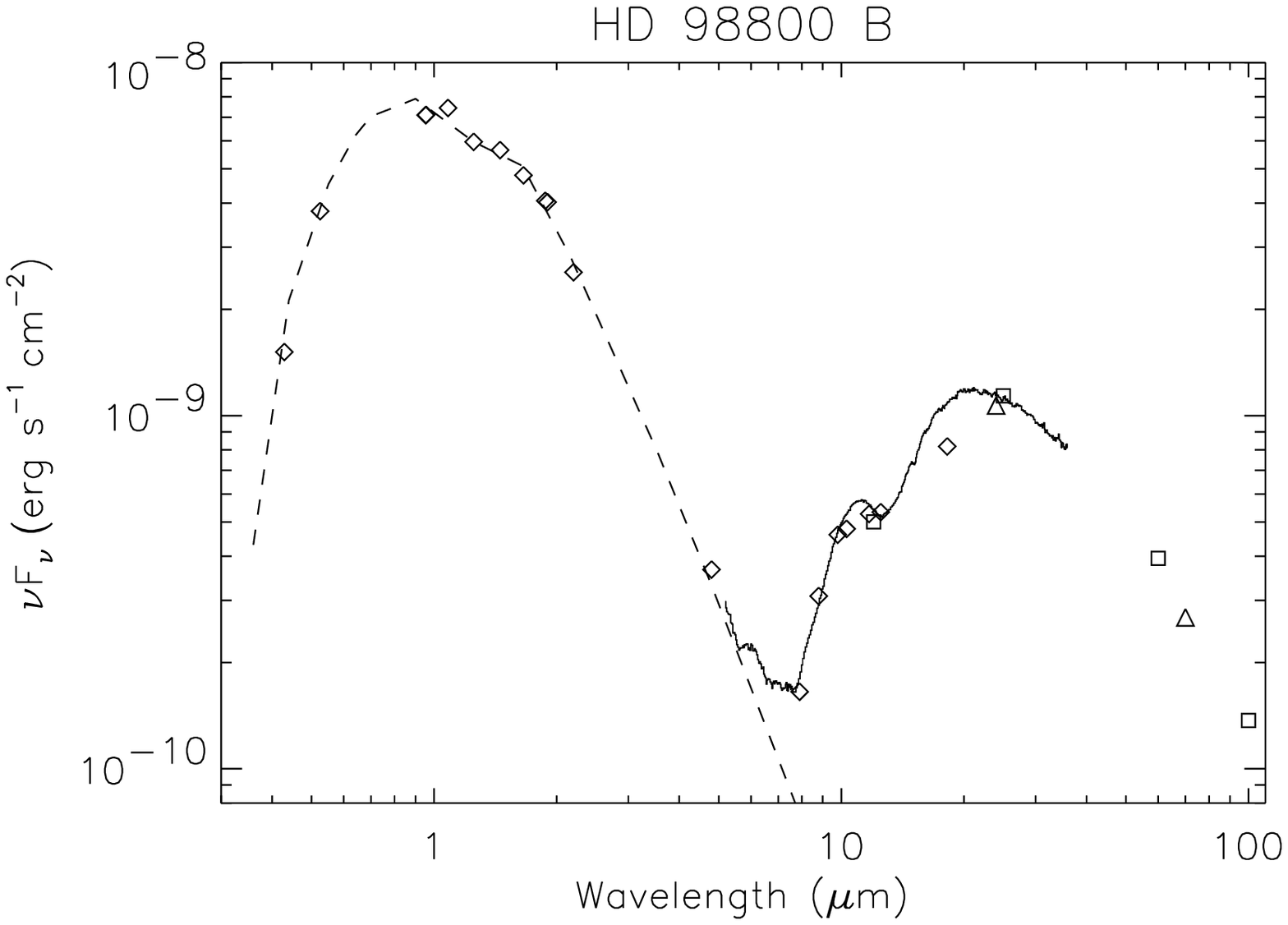}
\caption{SED of HD 98800 B, composed of optical to mid-IR photometry
compiled by \citet{prato01} ({\it diamonds}), {\it IRAS} fluxes ({\it squares}),
and the MIPS fluxes of \citet{low05} ({\it triangles}). The photosphere is also sketched 
in, based on the photospheric colors of a K5 star \citep{kenyon95} and 
normalized at the $J$-band flux given in \citet{prato01}. The IRS spectrum was
corrected for the contribution of the A component (see text for details). 
All fluxes were dereddened using $A_V$=0.44 and Mathis's reddening law 
\citep{mathis90}. \label{TWA4B_SED}}
\end{figure}

We estimated the mid-IR emission of HD 98800 A by using the photospheric 
colors of a K5 star \citep{kenyon95} 
normalized at the $J$-band flux of the A component given in \citet{prato01}.
This estimate is consistent with the measurements by \citet{prato01},
who determined that the contribution of A to the system flux decreases with 
wavelength (50\%, 44\%, 30\%, and 20\% at 2.2, 4.8, 7.9 $\mu$m, 
and 8.8 $\mu$m, respectively, and less than 10\% at longer wavelengths). 
The subtraction of the estimated emission of the A component from the IRS 
spectrum will affect both the flux level and shape of the spectrum particularly 
in the 5--8 $\mu$m wavelength range; given that we do not account for
atmospheric features, the resulting spectrum carries a higher uncertainty
in this wavelength region.

The SED of HD 98800 B is shown in Figure \ref{TWA4B_SED};
it was constructed with photometry of the B component compiled by \citet{prato01}, 
MIPS fluxes from \citet{low05}, {\it IRAS} fluxes, the IRS spectrum corrected for 
the contribution of A, and the photosphere of B (approximated using the colors of
a K5 star normalized at the $J$-band flux of B). 
The emission of B is photospheric out to about 5.5 $\mu$m; beyond that, the
infrared excess sets in rather sharply. The steep rise of the SED at 8 $\mu$m
is reminiscent of that observed in CoKu Tau/4, a transition disk in the Taurus
star-forming region with an inner disk hole of 10 AU \citep{forrest04,dalessio05}.
The SED peaks at 20 $\mu$m, then decreases as a single-temperature blackbody. 
The outer disk is likely truncated by component A, whose distance from B is almost 50 AU 
at closest approach (\citealt{torres95}; note that the orbit of B around A is very 
eccentric and seen almost edge-on). Thus, the disk around the spectroscopic
binary HD 98800 B seems to be truncated both at the inner and at the outer
disk edge. 

\subsection{Dust Composition}
\label{dust_section}

\citet{koerner00} inferred from narrow-band photometry in the 10-$\mu$m region
that the silicate emission feature of HD 98800 B is broader and more structured than
expected from amorphous silicates, indicating the presence of crystalline silicates. 
The IRS spectrum allows us to analyze the 10-$\mu$m silicate feature in greater
detail and to derive the dust species that generate the infrared excess emission. 

First, we subtracted the photospheric emission, constructed from the colors of a K5 star 
normalized at the $J$-band flux, from the dereddened IRS spectrum of HD 98800 B.
Next, given the large infrared excess and the presence of 10 and 18 $\mu$m silicate 
emission features, we fit the residuals with two components, an optically thick region
modeled as a blackbody, and a warmer, optically thin region of constant source function 
(also modeled as a blackbody):
\begin{equation} 
F_{\nu} =  \Omega_{thick} B_{\nu}(T_{thick}) + 
\Omega_{thin} B_{\nu}(T_{thin})(1-e^{-\tau_\nu}),
\end{equation}
where $T_{thick}$ is the blackbody temperature of the optically thick region, 
$T_{thin}$ is that of the optically thin region, $\tau_\nu$ is the wavelength-dependent 
optical depth of the optically thin cloud, and $\Omega_{thick}$ and $\Omega_{thin}$
are the solid angles of the emitting regions for the optically thick and thin components,
respectively. 

The optically thick component can be thought of as consisting of large (mm-sized) dust 
grains, which do not generate spectral features and whose opacities are independent of 
wavelength, thus creating a continuum component with a blackbody-type emission. 
Alternatively, a very large mass of small grains will generate blackbody-shaped,
optically thick emission. Thus, the detailed composition of the dust only enters in the 
optically thin component.

\begin{deluxetable*}{llllll}
\tabletypesize{\small}   
\tablewidth{\linewidth}
\tablecaption{Dust Component Fits of HD 98800 B \label{TWA4B_dust_comp}}
\tablehead{
\colhead{Model} & \colhead{Dust Species} & \colhead{$T_{thick}$} & 
\colhead{$T_{thin}$} & \colhead{$\Omega_{thick}$} & \colhead{${\chi}_{\nu}^2$} \\
 &  &  [K] & [K] & [10$^{-13}$ sr] & 
}
\startdata
a & 5 $\mu$m amorphous pyroxene & 155  & 373  & 1.39 & 18.6 \\
b & small amorphous carbon  &   &  \\
   & + 5 $\mu$m amorphous pyroxene & 151  & 291  & 1.47 & 12.1 \\
c & small amorphous carbon  &   &  \\
  & + 3 $\mu$m amorphous pyroxene & 153  & 298  & 1.43 & 11.4 \\
d & small amorphous carbon  &   &  \\ 
  & + 3 $\mu$m amorphous pyroxene  &  &  &  \\
  & + 3 $\mu$m amorphous olivine & 152  & 310  & 1.45 & 8.0 \\
\tableline
\enddata
\end{deluxetable*}

For optically thin emission, the flux becomes linearly dependent on the optical depth, 
which can be expressed as the sum of the optical depth of each dust component
$\tau_{\nu_i} = \kappa_{\nu_i} N_i$, where $\kappa_{\nu_i}$ and $N_i$ are the 
mass absorption coefficient and column density for each dust species, respectively. 
Since the mass of each dust species is derived via $m_i = N_i \Omega_{thin} d^2$, 
where $\Omega_{thin}$ is the solid angle subtended by the dust and $d$ is the distance 
to the system (assumed to be 47 pc), the total optical depth of the optically thin 
component can be expressed as 
\begin{equation}
\tau_{\nu} = \sum_{i=1}^N \frac{\kappa_{\nu_i} m_i}{\Omega_{thin} d^2},
\end{equation}
where the sum is carried out over the number $N$ of dust components.
Thus, the dust emission is calculated as
\begin{equation}
F_{\nu} =  \Omega_{thick} B_{\nu}(T_{thick}) + 
\frac{1}{d^2} B_{\nu}(T_{thin}) \sum_{i=1}^N \kappa_{\nu_i} m_i
\end{equation}
We solved the linear model for the best-fitting set of temperatures
$T_{thick}$ and $T_{thin}$, solid angle $\Omega_{thick}$, and dust masses $m_i$, 
starting with just one dust component and adding components as necessary to 
improve the fit.  

\begin{figure}
\centering
\includegraphics[angle=90,scale=0.33]{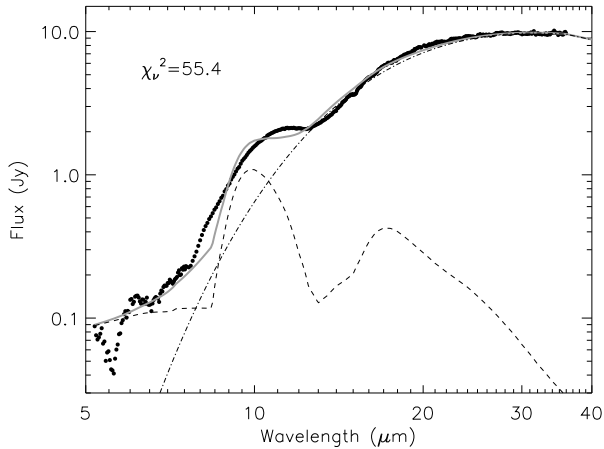}
\caption{Dust model fit, using ISM-type dust grains, for HD 98800 B ({\it thick gray line}), 
whose IRS spectrum is plotted after photosphere subtraction; note that the error 
bars for the IRS data points are generally smaller than the plotting symbols and are thus 
not shown. The model components are a blackbody at temperature $T_{BB}$=161 K 
for the optically thick part ({\it dash-dotted line}) and sub-$\mu$m amorphous olivine 
at T=662 K for the optically thin part ({\it dashed line}). The solid angle of the emitting 
area of the optically thick component is equal to 1.23 $\times$ 10$^{-13}$ sr. 
The large ${\chi}_{\nu}^2$ value of 55.4 (calculated between 6.5 $\mu$m 
and 1.3 mm) proves that this model is not a good fit. 
\label{TWA4B_poor_dust_fit}}
\end{figure}

For the optically thin dust species, we adopted amorphous carbon and silicates,
which are typical for the interstellar medium. We used optical constants from
\citet{draine84} for $\lambda <$ 7.5 $\mu$m and from \citet{dorschner95} 
for $\lambda \geq$ 7.5 $\mu$m (using power laws to extrapolate values from 
200 $\mu$m to 2 mm) for amorphous silicates of olivine (MgFeSiO$_4$) and 
pyroxene (Mg$_{0.8}$Fe$_{0.2}$SiO$_3$) composition, 
and optical constants from \citet{zubko96} for amorphous carbon. For sub-micron 
grains, we used a CDE2 shape distribution \citep{fabian01} to compute opacities, while
larger grains ($\gtrsim$ 1 $\mu$m) were adopted to be porous spheres with 
a 50\% volume fraction of vacuum. 

The first model attempts, using only sub-micron (i.e., ISM-like) olivine or pyroxene 
grains as the optically thin component, yielded poor fits (for an example, see Figure 
\ref{TWA4B_poor_dust_fit}); the emission at the long-wavelength side of both the 
10 and 18 $\mu$m silicate feature is underestimated, suggesting the presence of 
larger grains. Amorphous pyroxene represents a better fit to the observed peak positions 
of the 10 and 18 $\mu$m features than amorphous olivine, so it was assumed as the 
dominant large grain species.

\begin{figure*}
\centering
\includegraphics[angle=90,scale=0.75]{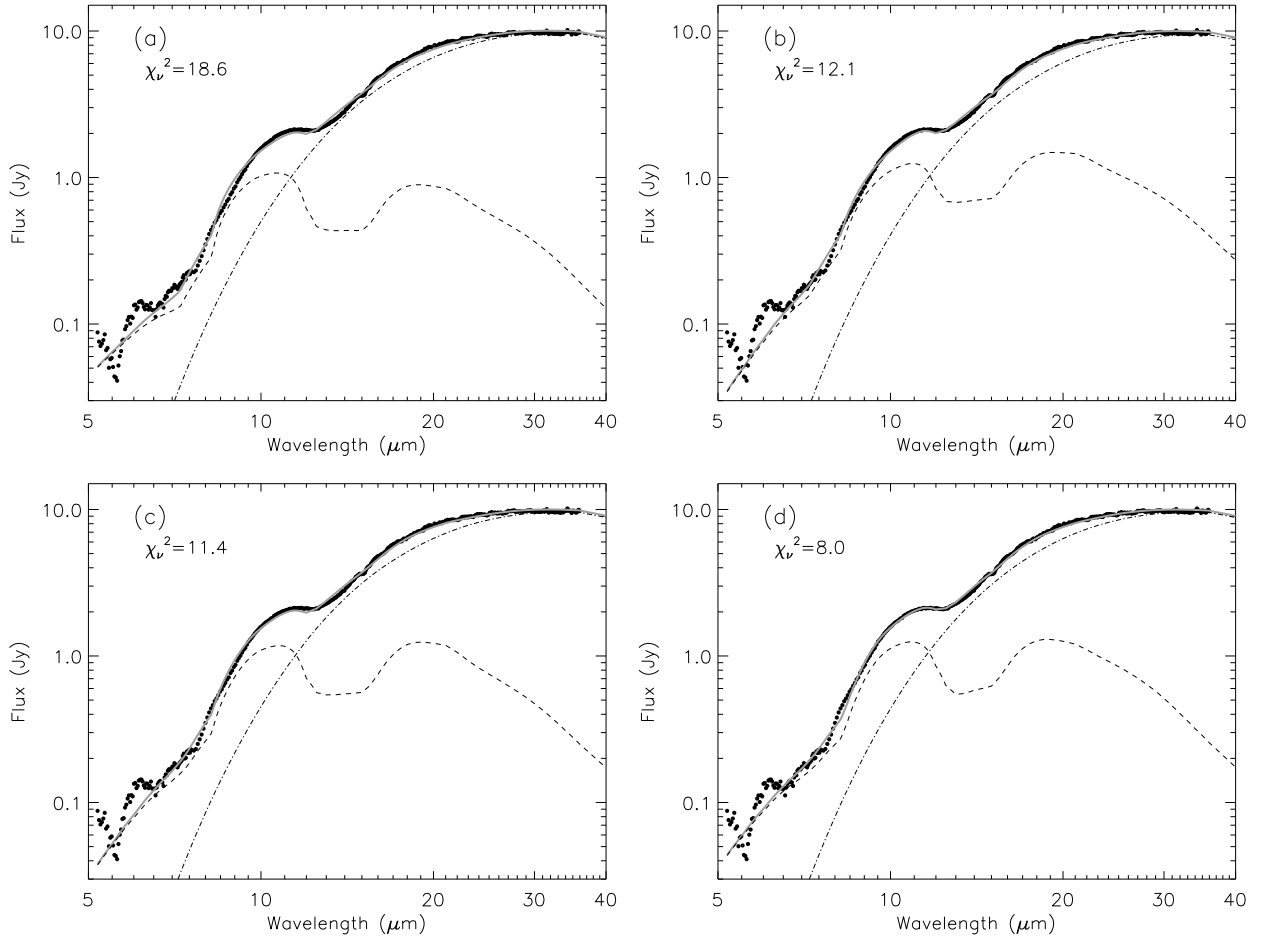}
\caption{Best dust model fits for HD 98800 B ({\it thick gray lines}); the IRS 
spectrum is plotted as in Figure \ref{TWA4B_poor_dust_fit}.
The ${\chi}_{\nu}^2$ values for each model are also shown in each panel.
The fits consist of blackbodies at different temperatures ($T_{BB}$) for the optically 
thick component ({\it dash-dotted lines}) and different dust species for the optically 
thin component ({\it dashed lines}): 
{\it (a)} $T_{BB}$=155 K, and 5-$\mu$m-sized amorphous pyroxene grains;
{\it (b)} $T_{BB}$=151 K, and small amorphous carbon and 5-$\mu$m-sized 
amorphous pyroxene grains;
{\it (c)} $T_{BB}$=153 K, and small amorphous carbon and 3-$\mu$m-sized 
amorphous pyroxene grains;
{\it (d)} $T_{BB}$=152 K and small amorphous carbon, 3-$\mu$m-sized 
amorphous pyroxene, and 3-$\mu$m-sized amorphous olivine grains.
\label{TWA4B_dust_fits}}
\end{figure*}

The best-fitting models, ordered by decreasing reduced ${\chi}^2$ values 
(${\chi}_{\nu}^2$), which were calculated from 6.5 $\mu$m to 1.3 mm, 
are shown in Figure \ref{TWA4B_dust_fits} with the residuals of HD 98800 B. 
The adopted model components and their parameters are displayed in 
Table \ref{TWA4B_dust_comp}. We note that most of the structure seen in the 5-6.5 
$\mu$m region of the residuals is likely due to photospheric features from the four stellar 
components of the system, which we did not account for in our photosphere subtraction 
process. Thus, we expect a mismatch between models and residuals in this part of the 
spectrum, and we did not include these data points in the computation of our 
${\chi}_{\nu}^2$ values. 

We first tried a model with an optically thick component at 155 K and a single 
optically thin component consisting of 5 $\mu$m amorphous pyroxene (Fig.\ 
\ref{TWA4B_dust_fits} a). This fit improved by adding sub-micron amorphous 
carbon to the dust mixture (Fig.\ \ref{TWA4B_dust_fits} b); in particular, the 
short-wavelength excess starting at about 6.5 $\mu$m is fit better. A slightly 
better fit of the silicate feature was achieved by using 3 $\mu$m amorphous 
pyroxene and small amorphous carbon as the optically thin component (Fig.\ 
\ref{TWA4B_dust_fits} c). Finally, the best fit was obtained by adding 3 $\mu$m 
amorphous olivine to the previous mixture and adopting an optically thick component 
at 152 K (Fig.\ \ref{TWA4B_dust_fits} d). 
For all of the models shown in Figure \ref{TWA4B_dust_fits}, the adopted solid angle lies 
in the (1.4-1.5) $\times$ 10$^{-13}$ sr range, and the temperature of the optically 
thin grains is higher than that of the optically thick component by about a factor of two 
(see Table \ref{TWA4B_dust_comp}).
Even though the differences between these four models are small (especially if
judged by eye), the best fit determined here for the optically thin disk region 
around HD 98800 B will be confirmed by our disk model in \S\ \ref{model_section}.

\begin{figure*}
\centering
\includegraphics[angle=90, scale=0.6]{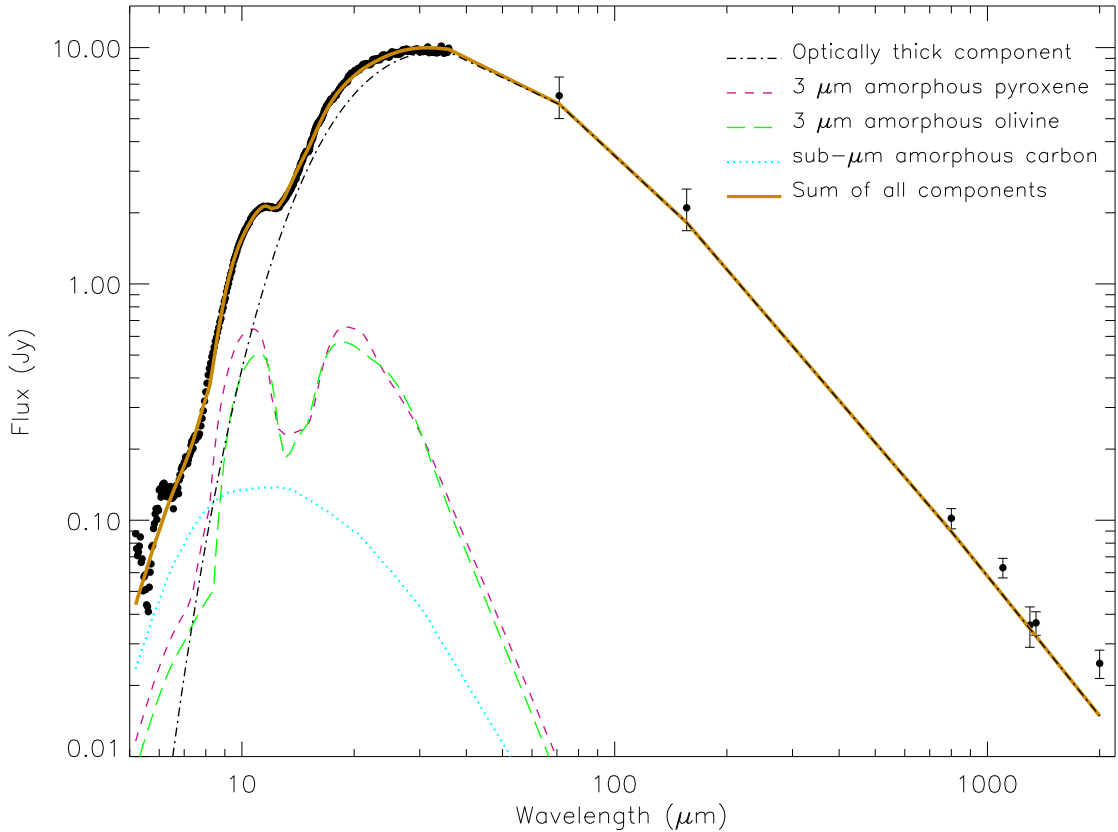}
\caption{The best-fitting dust model of HD 98800 B, plotted with the 
photosphere-subtracted IRS spectrum, long-wavelength photometry compiled
by \citet{prato01}, MIPS 70 and 160 $\mu$m fluxes \citep{low05}, and
SCUBA 1350 and 2000 $\mu$m fluxes \citep{sylvester01}. 
The various model components are listed in the top right quadrant of the figure.
\label{TWA4B_best_dust_fit}}
\end{figure*}

The best model fit and its components are shown in Figure \ref{TWA4B_best_dust_fit}
with the residuals of HD 98800 B over the mid-infrared to mm wavelength range.
The model fits the entire excess emission from 6.5 $\mu$m to 1.3 mm remarkably 
well. Table \ref{TWA4B_best_dust_model} lists the parameter values of this model 
together with their 1-$\sigma$ uncertainties, which were estimated from models lying 
within $\Delta {\chi}_{\nu}^2=1$ of the best-fitting model. Since the optically thick
component is more dominant and tightly constrained by all the measurements beyond
about 14 $\mu$m, the relative uncertainties for both $\Omega_{thick}$ and $T_{thick}$ 
are smaller than for the parameters describing the optically thin component.
We derive a combined mass for the optically thin dust grains of 4.7 $\times$ 10$^{-4}$ 
lunar masses. The mass of the optically thick component cannot be constrained, but it is 
likely several orders of magnitude larger than the mass of the smaller grains.

\begin{deluxetable}{ll}
\tabletypesize{\small}   
\tablewidth{\linewidth}
\tablecaption{Parameters and Uncertainties for the Best Dust Component Fit of 
HD 98800 B \label{TWA4B_best_dust_model}}
\tablehead{
\colhead{Parameter} & \colhead{Value} 
}
\startdata
$\Omega_{thick}$              & $1.46 \pm 0.02 \times 10^{-13}$ sr \\
$T_{thick}$                         & $151.8 \pm 0.4$ K \\
$T_{thin}$                          &  $310 \pm 2$ K \\
$m_{\mathrm{pyroxene}}$ & $2.3 \pm 0.2 \times 10^{-4}$ lunar masses \\
$m_{\mathrm{olivine}}$      & $1.7 \pm 0.2 \times 10^{-4}$ lunar masses \\
$m_{\mathrm{carbon}}$     & $6.9 \pm 0.8 \times 10^{-5}$ lunar masses \\
\tableline
\enddata
\tablecomments{
The mass of the moon is equal to $7.35 \times 10^{25}$ g.
}
\end{deluxetable}

The fact that dust grains larger than ISM-type amorphous grains (whose sizes are
$\lesssim$ 1 $\mu$m) constitute the bulk of the optically thin component indicates 
that dust growth must have occurred in the disk of HD 98800 B. 
The other two objects with substantial disks in the TW Hydrae association, TW Hya 
(K7 spectral type) and Hen 3-600 A (M3 spectral type), are dominated by smaller
grains and are characterized, in particular in the case of Hen 3-600 A, by a 
notable fraction of crystalline grains \citep{sargent06}. 
Our dust model fit to HD 98800 B excludes any sizable amount of small crystalline silicates,
suggesting that the larger width of the 10 $\mu$m silicate feature already noted by 
\citet{koerner00} should likely be attributed to grain growth alone.

\subsection{Disk Model}
\label{model_section}

In order to derive the structure of the circumstellar material of HD 98800 B, in
addition to its composition, we computed a model following the methods of 
\citet{dalessio05} and \citet{calvet05}. 
The stellar luminosity of the Ba+Bb pair was adopted to be 0.7 \Lsun, a value
obtained by integrating the photospheric fluxes of the unresolved B component,
and the effective temperature was assumed to be 4350 K, typical for a K5 star
\citep{kenyon95}.
The model has two components: an optically thick wall and an optically thin region, 
corresponding to the blackbody component and the optically thin dust grains,
respectively, from the dust component fit of \S\ \ref{dust_section}. 
Even though the model implicitly assumes the presence of gas in addition to the dust,
no assumptions regarding the gas were made, since it does not enter any of the 
calculations.

\begin{figure*}
\centering
\includegraphics[angle=90,scale=0.7]{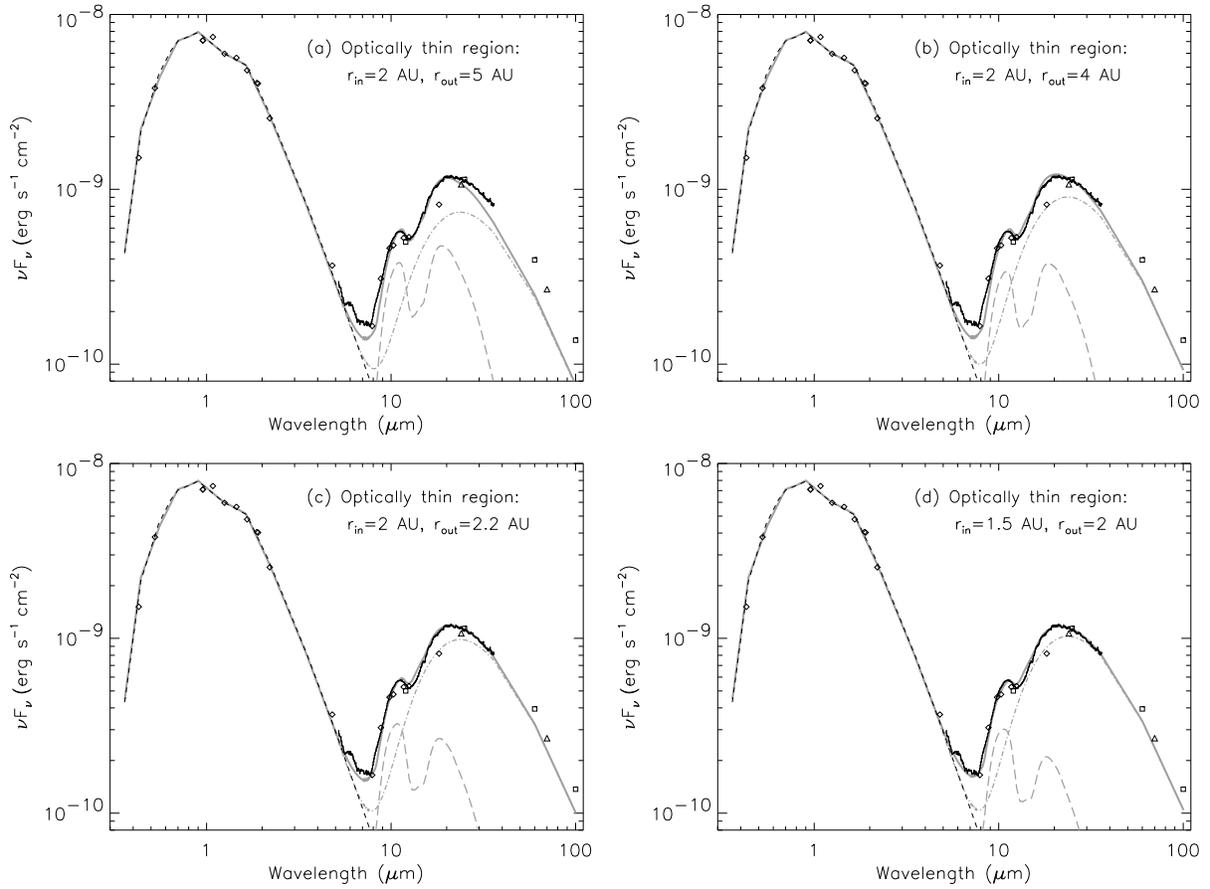}
\caption{Sequence of disk models for HD 98800 B ({\it thick gray lines}); the SED
is plotted as in Figure \ref{TWA4B_SED}. Also shown are the components of the 
disk models, the optically thick disk wall ({\it dash-dotted lines}) and the optically thin 
inner disk region ({\it long-dashed lines}). The four panels display different models
where the optically thin region consists of a ring between $r_{in}$ and $r_{out}$,
and the optically thick wall at 5.9 AU is scaled by means of the wall height to 
result in a good overall fit (see Table \ref{TWA4B_model_seq} for details). 
\label{TWA4B_disk_models}}
\end{figure*}

The wall emission was calculated assuming amorphous carbon and silicate grains with 
sizes between 1 and 3 mm. Since the grains are so large, their opacities are gray (i.e., 
independent of wavelength), and thus their composition does not affect the fit; their 
emission corresponds to that of a scaled blackbody. Therefore, this wall matches the 
optically thick component of \S\ \ref{dust_section}.
The composition of the optically thin component was adopted from \S\ \ref{dust_section},
namely sub-micron amorphous carbon, 3 $\mu$m-sized amorphous olivine and 3 
$\mu$m-sized amorphous pyroxene grains. 
The small amorphous carbon grains, which absorb stellar radiation more efficiently than the 
larger grains, are necessary to heat the dust and yield enough emission in the 6-8 $\mu$m 
range; the 3 $\mu$m-sized grains generate emission that provides a good fit to the silicate 
features at 10 and 20 $\mu$m.

\begin{deluxetable*}{llllllll}
\tabletypesize{\small}   
\tablewidth{\linewidth}
\tablecaption{Disk Model Fits of HD 98800 B \label{TWA4B_model_seq}}
\tablehead{
Model  &  \multicolumn{5}{c}{Optically thin component} & 
 \multicolumn{2}{c}{Optically thick component} \\
 & $r_{in}$ & $r_{out}$ & $\tau_{10}$ & $T(r_{in})$ & $T(r_{out})$ & 
$R_{wall}$ & $H_{wall}$ \\
 &  [AU] & [AU] & & [K] & [K] & [AU] & [AU] 
}
\startdata
a & 2 & 5 & 0.04 & 266 & 175 & 5.9 & 0.56 \\
b & 2 & 4 & 0.04 & 266 & 193 & 5.9 & 0.68 \\
c & 2 & 2.2 & 0.22 & 267 & 255 & 5.9 & 0.75 \\
d & 1.5 & 2 & 0.06 & 307 & 266 &5.9 & 0.75 \\
\tableline
\enddata
\tablecomments{
$\tau_{10}$ is used a scaling factor for the emission of the optically thin component:
$ F_{\nu} = \int_{r_{in}}^{r_{out}} B_{\nu}(T_{dust}) \kappa_{\nu}
\frac{\tau_{10}}{\kappa(10 {\mu}m)}2 \pi r dr$
}
\end{deluxetable*}

The results of the model calculations are shown in Figure \ref{TWA4B_disk_models} 
and in Table \ref{TWA4B_model_seq}. We show different model fits which result
from varying the contribution of the optically thin and thick components.
As in the previous section, the properties of the inner disk wall are well-constrained by
the flux measurements beyond about 14 $\mu$m; in order to obtain the correct
blackbody shape, it has to be placed at a distance of 5.9 AU from the B pair, where
the temperature in its upper layers (closest to the star) amounts to 140 K. 
Not surprisingly, this temperature is very similar to the one derived in \S\ 
\ref{dust_section} for the optically thick component.
The height of the wall above the midplane is treated as a scale factor for the
wall emission; it is adjusted such that the emission from the optically thick and thin
components yields the best possible fit. For the best-fit model (Figure 
\ref{TWA4B_disk_models} d), this height is 0.75 AU. 
The solid angle subtended by the wall amounts to 1.57 $\times$ 10$^{-13}$ sr
(assuming a distance of 47 pc and an inclination angle of 67\degr, as determined
by \citealt{boden05}), which is just somewhat larger than the value of 
1.46 $\times$ 10$^{-13}$ sr derived  from the best-fitting dust model in \S\ 
\ref{dust_section}. 

The optically thin component has to lie inside the region delimited by the optically 
thick wall in order to provide enough emission at wavelengths between 5 and 
20 $\mu$m; in addition, its innermost location and radial extent are fairly well 
constrained by the observed 10-20 $\mu$m flux ratio. 
As can be seen in Figure \ref{TWA4B_disk_models} a, an optically thin region 
extending from 2 to 5 AU will underestimate both the short- and long-wavelength 
excess emission; decreasing the outer radius results in a better match of the flux
beyond 20 $\mu$m (Fig.\ \ref{TWA4B_disk_models} b). Concentrating the
optically thin region close to 2 AU yields a very good fit (Fig.\ \ref{TWA4B_disk_models} 
c), while the best fit to the observations is obtained by an optically thin ring located 
between 1.5 and 2 AU (Fig.\ \ref{TWA4B_disk_models} d). The dust grains in this 
ring are at a temperature between 307 and 266 K, similar to the value found for the 
best-fitting dust model of \S\ \ref{dust_section}.

The wall radius of 5.9 AU we derive is comparable to the inner radius range of 
5.0 $\pm$ 2.5 AU determined by \citet{koerner00} from simple SED models;
both values are larger than the estimate of 2 AU derived by \citet{prato01} from
fits to their unresolved mid-infrared images. However, our result is consistent with
the fact that \citet{prato01} did not resolve the disk: the optically thin region 
is very compact and close to the binary, and even though at 12 $\mu$m the 
brightness of this region and that of the wall are comparable, the large inclination 
of the wall might have prevented its direct detection.

\begin{figure}
\centering
\includegraphics[scale=0.5]{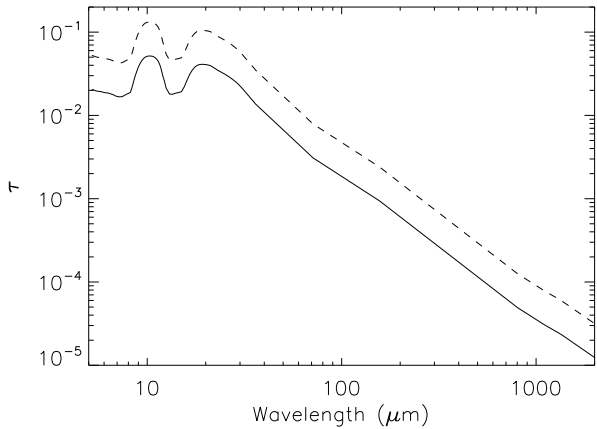}
\caption{Optical depth as a function of wavelength for the optically thin component 
of the best-fitting dust model of HD 98800 B. The solid line shows the vertical optical 
depth, while the dashed line represents an upper limit to the optical depth along the line 
of sight. \label{TWA4B_dust_taus}}
\end{figure}

The optical depth of the optically thin component is displayed in Figure 
\ref{TWA4B_dust_taus}; it was computed using the best-fitting dust component
model from \S\ \ref{dust_section} and adopting a ring between 1.5 and 2 AU
to determine $\Omega_{thin}$, which amounts to 5.85 $\times 10^{-14}$ sr.
Shown in the figure are the vertical optical depth ({\it solid line}) and the optical
depth along the line of sight ({\it dashed line}), assuming an inclination angle of 
67\degr. The latter quantity is actually an upper limit to the optical depth along
our line of sight, since the thickness of the optically thin region is not known. Thus,
since $\tau_\nu$ lies between 0.02 and at most 0.13 over the mid-infrared 
range and decreases steadily towards longer wavelengths, the inner disk region 
will appear as optically thin from infrared to millimeter wavelengths.

\begin{figure}
\plotone{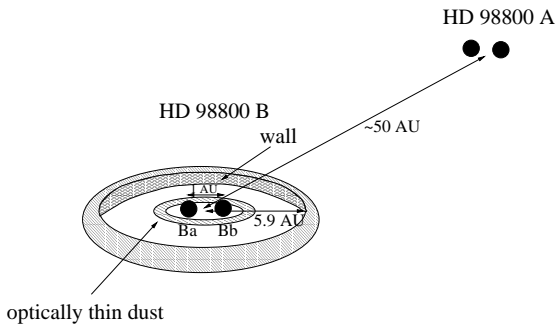}
\caption{Sketch of the HD 98800 system. {\it Note:} Distances and sizes 
of the disk components around HD 98800 B are not plotted to scale. 
\label{HD98800_sketch}}
\end{figure}

A sketch of the HD 98800 system is shown in Figure~\ref{HD98800_sketch}. 
There are two rings of material around the B component: an optically thick
wall at 5.9 AU with a small radial extent, and an optically thin inner region
between 1.5 and 2 AU, just outside the binary orbit. We will discuss the
implications of this peculiar structure in the next section.

\section{Discussion and Conclusions}
\label{disc_concl}

The mid-infrared spectrum of HD 98800 B reveals that it is surrounded by a transition
disk whose infrared excess emission starts at about 5.5 $\mu$m and decreases
beyond 20 $\mu$m. Its structure is somewhat reminiscent of that of the transition
disks TW Hya \citep{calvet02} and GM Aur \citep{calvet05}, which also harbor
optically thin regions inside an outer, optically thick disk. However, as opposed to the
these two objects, HD 98800 is a multiple system: the Ba and Bb stars form a
close ($\sim$ 1 AU) eccentric ($e=0.78$; \citealt{boden05}) binary system,
and in addition the A pair comes as close as 50 AU to the B pair, with the orbital planes 
of the two binaries inclined by about 30\degr\ with respect to each other 
\citep{torres95,boden05}. Thus, gravitational perturbations likely play a role in 
the confinement and distribution of the dust around HD 98800 B.

A circumbinary disk around an eccentric binary is expected to be tidally truncated
at the inner disk edge \citep{artymowicz94}. If the eccentricity is large,
several higher-order resonances play a role in truncating the circumbinary disk 
by exerting tidal torques further out in the disk; thus, gap sizes of $\sim$ 
3.5$a$ (where $a$ is the semimajor axis of the binary) can be explained
\citep{artymowicz94}. However, these gravitational interactions apply to
a gaseous disk, and the amount of gas in HD 98800 appears to be very small: 
the 1-$\sigma$ upper limit placed by \citet{dent05} on the sub-mm 
$^{12}$CO J=3-2 line indicates an upper limit of $\sim$ 4 $\times$ 10$^{-4}$ 
Jupiter masses (which is less than 1/24 of the gas mass of TW Hya), assuming 
optically thin $^{12}$CO emission. Some warm gas could still be present in
the inner disk regions, since the sub-mm $^{12}$CO line traces cold molecular 
gas in outer disk regions (where temperatures are below $\sim$ 100 K), but 
given the lack of accretion signatures \citep{soderblom96,webb99}, the amount 
of gas in the inner disk should be small.

In the absence of gas, the truncation of the outer disk at 5.9 AU could still be explained 
by the effect of resonances. Using the empirical formula determined by \citet{holman99}, 
who studied stable particle orbits around an eccentric binary, the smallest, stable orbit 
around HD 98800 B (which has a semimajor axis $a=0.98$ AU, eccentricity $e=0.78$, 
and mass ratio $\mu=M_2/(M_1+M_2)=$ 0.45; \citealt{boden05}) is at 4.06 AU, 
somewhat smaller than the result we obtained by fitting a disk model to the SED. 
However, \citet{holman99} note that the transition between stable and unstable 
orbits is likely not sharp due to the effects of overlapping mean motion resonances. 
They also found indications that, over time, the stable region moves further out 
for high binary eccentricities. 

While a tidally truncated disk at $\gtrsim$ 4 AU could explain the location of the disk
wall, it would not be able to account for the presence of optically thin dust grains at 1.5-2 AU. 
The most likely interpretation for the unusual structure of the circumstellar material of 
HD 98800 B, also considering the probable scarcity of gas, is that the disk is already 
at the debris disks stage, when dust is its main constituent and is generated 
by collisions of larger bodies. The dust would be second-generation dust and not 
primordial material that survived for 10 Myr. 
The sub-micron carbon and 3-$\mu$m-sized silicate grains in the disk would be 
replenished in planetesimal collisions, which might occur in the optically thick ring 
at $\sim$ 6 AU. Due to Poynting-Robertson (PR) drag, the larger dust grains 
would spiral in from 6 AU towards the binary on a timescale of a few 10$^5$ 
years; since $t_{PR} \propto a D^2/L_{\ast}$ \citep{burns79}, where 
$a$ is the grain size, $D=6$ AU and $L_{\ast}=0.7$ \Lsun, smaller grains 
would migrate even faster. Since HD 98800 is about 10 Myr old, the optically 
thin, inner ring must be continuously replenished, assuming it is a long-lived structure.

If the inner dust ring is explained by PR drag, then the absence of grains from 2 to 
5.9 AU is puzzling; drag forces acting on dust grains should distribute the dust uniformly 
inside the radius at which the dust grains are created. 
A possible explanation for the observed gap could be a planet that formed just outside the 
unstable region, i.e., close to the inner disk wall, and that is temporarily holding up dust 
grains (that were able to drift inward) at one of its inner mean motion resonances
\citep{liou97,moro-martin05}.
In the case of TW Hya, a planet was thought to be responsible for clearing out the inner 
disk \citep{calvet02}, so it is conceivable that a planet also formed in the roughly coeval 
HD 98800 system.
On the other hand, this system is probably governed by complex dynamics due to the 
presence of four stellar components, implying overlapping resonances and variable
gravitational perturbations. 

Given that the evidence supports that the disk around HD 98800 B is rather a 
debris than a protoplanetary disk, the presence of an optically thick dust component 
and the large infrared excess (L$_{IR}$/L$_{bol}=$17\%) of HD 98800 B seem
unusual, but they might be explained by the gravitational perturbations of the Aa+Ab pair.
This type of perturbation can pump up eccentricities and inclinations of particles,
and cause particles to be trapped in mean motion resonances, as was likely the case
for Kuiper Belt objects under the influence of the giant planets and possibly a close
encounter by a passing, nearby star \citep[e.g.,][]{duncan95,ida00,gladman05}. 
Periodic stirring of planetesimals in the outer disk around HD 98800 B by the A pair could 
be responsible for generating copious amounts of dust \citep[see][]{kenyon02}.
HD 98800 B is thus a unique type of debris disk, whose infrared excess is elevated
to levels comparable to that of protoplanetary disks due to the particular configuration
of the four components in this system, resulting in gravitational perturbations that
prevent the dust from settling into a flat disk.

Even though HD 98800 appears to be a very dynamical system, it is unlikely to be in 
a short-lived transitional stage with ongoing clearing processes, in which the outer disk
is being progressively eroded.
HD 98800 B belongs to a similar class of transition disks as St 34 and Hen 3-600 A:
a tight binary is responsible for tidal and resonant interactions with the disk, thus creating 
a stable, tidally truncated circumbinary disk. In addition, collisions between planetesimals
in the outer disk of HD 98800 B cause a collisional cascade of smaller grains, which then
migrate towards the central binary. This outer disk is likely truncated due to the presence 
of the other component, as is the case with Hen 3-600 A \citep{jayawardhana99a}.
However, St 34 and Hen 3-600 A are still accreting material, albeit at low levels 
(2.5 $\times$ 10$^{-10}$ $M_{\odot}$ yr$^{-1}$ for St 34, $\sim$ 5 
$\times$ 10$^{-11}$ $M_{\odot}$ yr$^{-1}$ for Hen 3-600 A; 
\citealt{white05,muzerolle00}), while HD 98800 B is probably not accreting
any more.
Thus, St 34 and Hen 3-600 A are likely surrounded by evolved protoplanetary disks, 
while the disk around HD 98800 seems to have evolved even further. The similarities 
and differences between the HD 98800 and Hen 3-600 systems could increase our
understanding of disk evolution; the larger separation between Hen 3-600 Aa+Ab 
and B and the fact that all three stars are of later spectral type might play a role 
in the longer survival of primordial disk material around this system.

The analysis presented in this paper suggests that transition disks comprise a 
varied group of objects, where different processes are responsible for 
creating a ``transition disk'' appearance, and where the timescales involved
can vary substantially. Increasing the sample of transition disks from various
star-forming environments and at different ages will shed light on the processes
sculpting the disks. \\

\acknowledgments
We thank an anonymous referee whose comments led to a substantial improvement
of this paper and a better understanding of this object.
This work is based on observations made with the {\it Spitzer Space Telescope}, 
which is operated by the Jet Propulsion Laboratory, California Institute of Technology, 
under NASA contract 1407. Support for this work was provided by NASA through 
contract number 1257184 issued by JPL/Caltech. E.F. was supported by a NASA
Postdoctoral Program Fellowship, administered by Oak Ridge Associated Universities
through a contract with NASA. N.C. and L.H. acknowledge support
from NASA grants NAG5-13210 and NAG5-9670, and STScI grant AR-09524.01-A.
P.D. acknowledges grants from PAPIIT, UNAM and CONACyT, M\'exico.
This publication made use of NASA's Astrophysics Data System Abstract Service,
and of data products from the Two Micron All Sky Survey, which is a joint project of 
the University of Massachusetts and the Infrared Processing and Analysis Center/California 
Institute of Technology, funded by NASA and the NSF.

\end{document}